# Time-resolved spectroscopy of exciton states in single crystals, single crystalline films and powders of $YAlO_3$ and $YAlO_3$:Ce

**V. Babin[1], L. Grigorjeva[2], I. Kondakova[3], T. Kärner[1], V.V. Laguta[3,4], M. Nikl[4], K. Smits[2], S. Zazubovich[1], Yu. Zorenko[5]**

[1] *Institute of Physics, University of Tartu, Riia 142, 51014 Tartu, Estonia*

[2] *Institute of Solid State Physics, University of Latvia, Kengaraga 8, Riga, Latvia, LV-1063*

[3] *Institute for Problems of Material Science NAS Ukraine, Krjijanovskogo 3, 03142 Kiev, Ukraine*

[4] *Institute of Physics AS CR, Cukrovarnicka 10, 162 53 Prague, Czech Republic*

[5] *Ivan Franko National University of Lviv, Gen. Tarnavsky 107, 79017 Lviv, Ukraine*

**Abstract** Luminescence characteristics of single crystals (SC), single crystalline films (SCF), powders and ceramics of $YAlO_3$ and $YAlO_3$:Ce have been studied at 4.2-300 K under photoexcitation in the 4-20 eV energy range and X-ray excitation. The origin and structure of defects responsible for various exciton-related emission and excitation bands have been identified. The ~5.6 eV emission of $YAlO_3$ SCF is ascribed to the self-trapped excitons. In $YAlO_3$ SC, the dominating 5.63 eV and 4.12 eV emissions are ascribed to the excitons localized at the isolated antisite defect $Y^{3+}_{Al}$ and at the $Y^{3+}_{Al}$ defect associated with the nearest-neighbouring oxygen vacancy, respectively. Thermally stimulated release of electrons, trapped at these defects, takes place at 200 K and 280 K, respectively. The formation energies of various $Y^{3+}_{Al}$-related defects are calculated. The presence of $Y_{Al}$ antisite-related defects is confirmed by NMR measurements. The influence of various intrinsic and impurity defects on the luminescence characteristics of $Ce^{3+}$ centers is clarified.

## 1. Introduction

Single crystals (SC) of $Ce^{3+}$-doped yttrium-aluminum perovskite $YAlO_3$:Ce (YAP:Ce) were found to have favorable scintillation characteristics especially for medical applications [1, 2] which stimulated intense studies of this material. Luminescence characteristics of $Ce^{3+}$ centers, as well as energy transfer and defects creation processes in YAP:Ce SC were studied in [1-6]. The crystals grown at high temperatures from the melt contain antisite (e.g., $Y^{3+}_{Al}$-, $Ce^{3+}_{Al}$-type) and vacancy-related defects which can negatively influence the scintillation characteristics of YAP:Ce SC. The concentration of these defects is strongly suppressed in YAP single crystalline films (SCF) prepared by the low-temperature Liquid Phase Epitaxy (LPE) method [7]. However, at the preparation of the SCF with the use of the PbO-containing flux, lead ions are introduced into the SCF. As a result, various lead-related centers appear. Their luminescence characteristics were studied in [8-10].

The luminescence of undoped YAP SC was studied for more than thirty years (see, e.g., [7-9, 11-17] and references therein). Two emission bands observed at 5.9 eV (FWHM=0.7 eV) and 4.2 eV (FWHM=0.9 eV) under excitation in the exciton absorption region were ascribed to the self-trapped excitons (STE) and recombination processes, respectively [12-16]. In [7-9, 17], the role of antisite defects of the type of $Y^{3+}_{Al}$ in the formation of intrinsic luminescence centers was considered. After that, the highest-energy intrinsic emission of YAP SC was concluded to arise from the $Y^{3+}_{Al}$ defect itself [7], or the exciton localized near the $Y^{3+}_{Al}$ defect [8-9], or the STE [17], while the emission bands located in the 3.0-4.5 eV range, from the excitons localized near various vacancy-related defects [17].

Thus, in SC and SCF of YAP, several types of intrinsic defects can exist. This reduces the efficiency of energy transfer from the host lattice to a luminescence center. Photo-thermally stimulated disintegration of various exciton states into electron and hole centers and their subsequent tunneling recombination can result in the appearance of slow luminescence decay components. The overlap of intrinsic emission bands with the absorption bands of $Ce^{3+}$ can allow an effective energy transfer from the defect-related centers to $Ce^{3+}$ ions which negatively influences the scintillation characteristics of $Ce^{3+}$-doped YAP.

However, an origin of the defect-related states as well as the processes responsible for the luminescence of YAP are not still clear. In different papers, different data are reported on the positions of the exciton, defect-related and impurity-related excitation bands in the exciton absorption region (see, e.g., Table 1). Therefore, their detailed comparison is not always possible. In the present work, the characteristics of various emission bands measured for the same undoped and $Ce^{3+}$-doped YAP SC and YAP SCF samples at the same experimental conditions are compared. The aim of the work was to clarify the origin and structure of defects responsible for the unrelaxed (denoted further as perturbed) and the relaxed (localized) exciton states in the SC and SCF of YAP and to obtain an information on the processes of energy transfer from the defect-related centers to the luminescence $Ce^{3+}$ centers.

## 2. Experimental procedure

Single crystals of undoped and $Ce^{3+}$-doped YAP were grown in CRYTUR Ltd., Turnov, Czech Republic, by the Czochralski method in a molybdenum crucible and studied in [3, 6]. Single crystalline films of undoped YAP and YAP:Ce were prepared in Lviv University, Ukraine, by the LPE method from the PbO-$B_2O_3$ flux in a platinum crucible and investigated in [10]. The nanopowders of YAP and YAP:Ce with the size of particles of about 27 nm were synthesized by the coprepitation method. The ceramics were obtained by the heating of the mentioned powders at 1000 $^{o}$C.

Luminescence characteristics were studied at 10 K under excitation by the synchrotron radiation at SUPERLUMI station (HASYLAB at DESY, Hamburg, Germany) with the time windows 1-9 ns and 96-149 ns. The excitation spectra of different emissions were measured in the 4-20 eV energy range at exactly the same conditions and corrected for the spectral distribution of the excitation light. The emission spectra were measured with the use of the monochromator ARC 0.3 m Spectra Pro300i, detected with the photomultiplier Hamamatsu R6358P and not corrected. The experiments at low temperatures were carried out with the use of a close-circle refrigerator. The X-ray luminescence spectra were measured at 80 K (40 kV, 10 mA) using Andor Sharock B-303i spectrograph with CCD camera (Andor DU-401A-BV).

Thermally stimulated luminescence (TSL) glow curves were measured with the heating rate 5 K/min after X-ray irradiation of a sample at 70 K for 30 min (40 kV, 15 mA). The sample located in an Oxford Instruments OptistatCF (Continuous Flow) helium cryostat was cooled and heated by He exchange gas, the heating rate was controlled with a ITC5205 temperature controller. The luminescence was registered using a Hamamatsu H8259-01 photon counting head.

## 3. Results and discussion

In YAP SCF, all the intrinsic and defect- or impurity-related emissions are most effectively excited in the exciton absorption region. The emissions of YAP SC are effectively excited in the band-to-band transitions region as well. The positions of the excitation bands in the exciton region observed for various exciton emission bands as well as for the emission of $Ce^{3+}$ centers in YAP SC and YAP SCF, and single and dimer $Pb^{2+}$-based centers in YAP SCF are shown in Table 1. It is evident that they are characteristic for each emission band studied.

### *3.1. Self-trapped and localized exciton luminescence in YAP single crystalline films*

As it was mentioned above, the number of the antisite and vacancy-related defects is strongly suppressed in SCF as compared with SC. Therefore, in the emission spectrum of YAP SCF, the STE band should be the dominating one. In the uncorrected emission spectrum of YAP SCF, measured at 10 K under excitation in the 8.40-7.95 eV range, the most intense band peaking at ~5.6 eV is observed (Fig. 1a, curves 1-3) which should arise from the STE. The excitation band of this emission is located at about 7.83 eV (Fig. 2a) (see Table 1). The band gap energy in YAP is $E_g$=8.8 eV [18, 19].

Besides the STE emission, a complex band is observed in YAP SCF, consisting of at least three components with the maxima at 4.16 eV, 3.63 eV, and 3.15 eV (Fig. 1a, curve 4). Two later bands arise from the single and dimer $Pb^{2+}$-based centers [10]. A wide 4.16 eV emission, observed in [8] at 4.19 eV and ascribed to an exciton localized near a $Pb^{2+}$ ion, is excited around 7.6 eV (Fig. 2b). The origin of the analogous (4.3 eV) emission in LuAG SCF was studied in [20]. By analogy with [20], we suggest that the 4.16 eV emission arises from an exciton localized near a $Pt^{4+}$ ion whose presence is caused by the use of a platinum crucible at the SCF preparation. In the excitation spectrum of the 3.63 eV emission of the single $Pb^{2+}$-based centers, the 7.66 eV band is observed (Fig. 3a). In the excitation spectrum of the 3.15 eV emission of the dimer $\{Pb^{2+}-V_O-Pb^{2+}\}$ centers, the band located at 7.74 eV is present (Fig. 3b). These bands can arise from the excitons perturbed by the single and the dimer $Pb^{2+}$-related centers, respectively. The appearance of the inner-centre 3.63 eV and 3.15 eV emissions under excitation in the exciton 7.66 eV and 7.74 eV bands means that the excitation energy is effectively transferred to the corresponding lead centers. Different positions of the excitation bands of the 4.16 eV emission and the 3.63 eV, 3.15 eV emissions (Table 1) confirm our conclusion that the 4.16 eV emission is not connected with the $Pb^{2+}$-related centers.

### *3.2. Intrinsic luminescence of YAP single crystals*

In the uncorrected emission spectrum of YAP SC, the 5.63 eV and 4.12 eV emissions are observed (Fig. 1b). As it was mentioned in the Introduction, different interpretations were given in [7-9, 13-17] for the highest-energy intrinsic emission band. As in SC, containing a lot of antisite and vacancy-related defects, the STE emission should be strongly suppressed, the 5.63 eV emission is surely connected with the antisite defect. In [17], a weak 5.28 eV emission was ascribed to $ex^0Y^{3+}_{Al}$. However, this emission is practically absent under excitation in the exciton absorption region (Fig. 1b). This is evident also from the complete coincidence of the excitation spectra shown in Figs. 4a and 4b. The relative intensity of the 5.28 eV emission is much larger under X-ray excitation. By analogy with the ~4.3 eV emission of LuAG [20], we suggest that the 5.28 eV emission arises from recombination processes.

The 4.2 eV emission found in [12] has been ascribed to the exciton-based center [15, 16, 21]. This emission was observed also in the thermally stimulated luminescence (TSL) spectra of undoped YAP SC [13, 16, 22] and ascribed in [22] to an exciton localized around a single oxygen vacancy. However, according to [23], oxygen vacancies in YAP SC are mainly associated with the antisite $Y^{3+}_{Al}$ defects.

The results obtained indicate that the characteristics of the 5.63 eV and 4.12 eV emissions in YAP SC differ from the characteristics of the STE emission in YAP SCF. In the YAP SCF, these emissions are absent [8]. We conclude that both the 5.63 eV and the 4.12 eV emission arise from the excitons localized near $Y^{3+}_{Al}$-related defects. As the positions of their lowest-energy excitation bands are different: the 5.63 eV emission is excited at ~7.9 eV (Fig. 5a), while the 4.12 eV emission, at 7.13, 7.7 eV (Fig. 5b), it means that the defects responsible for these emissions are different as well. One can suggest that these emissions arise from an exciton localized near a single antisite defect: $ex^0Y^{3+}_{Al}$ and from an exciton localized near the $Y^{3+}_{Al}$ defect associated with an oxygen vacancy $V_O$: $ex^0\{Y^{3+}_{Al}$-$V_O\}$. Indeed, the defects of the type of $\{Y^{2+}_{Al}$-$V_O\}$ were recently detected by the ESR method in the irradiated YAP crystals [23]. Their emission band is located at 2.45 eV [24]. The origin of this emission is confirmed by the comparison of the X-ray excited luminescence spectra of YAP SC with those of YAP nanopowder and ceramics prepared at the temperatures below 1000 C. In the latter materials, unlike YAP SC, the emission bands in the 4.0-6.0 eV energy range are completely absent (Fig. 6) indicating to the absence of antisite defects. In ceramics and nanopowders, the 2.45 eV emission is absent as well. Unlike YAP SC, in these materials the centers with the 2.45 eV photoluminescence are not created by the X-ray irradiation.

Thermal stability of the $ex^0\{Y^{3+}_{Al}$-$V_O\}$ excitons should be higher as compared with the $ex^0Y^{3+}_{Al}$ excitons. Under 8.3-8.4 eV excitation, the 5.63 eV emission intensity is constant up to 150 K and then decreases twice at about 220 K [3, 12]. The reduction of the 5.63 eV emission intensity is accompanied with the enhancement of the 4.12 eV emission [3]. In YAP:Ce SC, the TSL peaks are observed at the temperatures (~200 K and ~280 K, Fig. 7) corresponding to the thermal quenching of the 5.63 eV and 4.12 eV emissions, respectively. However, these peaks are absent in the above-mentioned YAP ceramics and nanopowders (Fig. 7). Therefore, we conclude that they arise from the recombination with the hole $Ce^{4+}$ centers of electrons, thermally released from the $Y^{3+}_{Al}$ and $\{Y^{3+}_{Al}$-$V_O\}$ traps, respectively.

The assumption about effective formation of the $\{Y^{3+}_{Al}$-$V_O\}$-type defects in YAP SC is supported also by the comparison of the $Y^{3+}_{Al}$ and $\{Y^{3+}_{Al}$-$V_O\}$ defects formation energies.

### *3.3. Calculation of the $Y^{3+}_{Al}$, $\{Y^{3+}_{Al}$-$V_O\}$, and $\{Y^{3+}_{Al}$-$Ce^{3+}\}$ defects formation energies*

*Ab-initio* computations allow determination the total energies as a function of ionic coordinates and types of ions, and thus provide independent information about the microscopic structure of a defect and clarify some models of defects.

The density functional theory (DFT) calculations were performed by us within the local spin-density approximation (LSDA). We have used the all-electron full-potential local-orbital (FPLO)_ code _version 7.00–28 [25]. In our scalar relativistic calculations, the exchange and correlation potential of Perdew and Wang [26] was employed. The **k**-mesh gives the subdivision of the Brillouin zone along the three axes from which the *k*-space integration mesh is constructed. We have used the default value 12x12x12 for pure $YAlO_3$ and 6x6x6 for supercells with defects. The defect energy was calculated using 20-atom $P_{nma}$-symmetry cell , where the lattice parameters were taken from x-ray diffraction data [27].

Using the supercell calculations we obtained the formation energy of antisite defect by comparing the ground state total energy of perfect cell with that of cells with single Y-Al interchanges. The structural parameters used are summarized in Table 2. As can be seen from the table that we obtained using the LDA calculations, the relaxed structure is close to the experimentally found [27] and is consistent with published earlier density functional results [28] .

The calculated energies are summarised in Table 3. We found the antisite defect formation energies in the presence of the nearest-neighbouring oxygen vacancy as well as in the presence of $Ce^{3+}$ impurity ion. The formation of antisite defect is preferable in cases where there is an oxygen vacancy near, which settles close to the direction Al - $V_O$ - Al. The formation energy of such type defect is by 25 % lower as compared with the energy of the antisite ion with vacancy $V_O$ in more far environment.

We have also calculated the total energy for supercell in which $Ce^{3+}$ ion replaced Y ion and found that the antisite defect formation is facilitated when $Ce^{3+}$ occupies the nearest $Y^{3+}$ site (see Table 3).

Our calculations have shown also that the ground state energy of the $Ce^{3+}$-doped YAP lattice with the oxygen vacancy near $Ce^{3+}$ is lower on 0.25 eV than the energy of the lattice in which the oxygen vacancy are located far from the $Ce^{3+}$ ion.

In the case when the $Y^{3+}$ substitutes for the Al nearby the oxygen vacancy, the $Y^{3+}$ ion must be shifted towards the vacancy because $Y^{3+}$ ion is much larger than $Al^{3+}$ ion. The total energy dependence on the displacement of $Y^{3+}$ antisite ion is presented in Fig. 8. One can see that the calculation predicts the $Y^{3+}$ ion displacement from the Al position in the direction towards the oxygen vacancy by the distance about 0.34 Å.

### *3.4. The excitons perturbed by $Ce^{3+}$-related centers in SC and SCF of YAP:Ce*

At 4.2 K, the emission spectrum of $Ce^{3+}$ centers in YAP:Ce consists of two components peaking at 3.30 eV and 3.55 eV (see, e.g., [3, 6]). In SC and SCF of YAP:Ce, the excitation spectra of the $Ce^{3+}$ emission in the exciton absorption region are considerably different (see also [7, 9]). The main exciton band is located at ~7.7 eV in SC (Fig. 9a), but at about 7.9 eV, in SCF (Fig. 9b). These bands are most probably arising from an exciton perturbed by a $Ce^{3+}$-related defect. As the main difference between SCF and SC is in the absence of antisite defects in SCF, we assume that the 7.9 eV band arises from an exciton perturbed by a single $Ce^{3+}$ ion, while the band around 7.7 eV, from an exciton perturbed by a $Ce^{3+}$ ion associated with the antisite defect $Y^{3+}_{Al}$ (see Table 1). This suggestion is confirmed by the fact, that in LuYAP:Ce, where the content of antisite defects should be larger [29], the above-mentioned difference in the excitation spectra in SC and SCF is also larger [9]. The presence of non-equivalent $Ce^{3+}$ centers in YAP:Ce and LuYAP:Ce has been noticed also in [6, 30].

The comparison of the excitation spectra of the slow and fast decay components of the $Ce^{3+}$ emission in the exciton region (Fig. 9) indicates that in YAP SC, unlike YAP SCF, a competition exists between the processes responsible for these components (see also [7-9]). Indeed, the excitation spectrum maximum of the fast component coincides with the excitation spectrum minimum of the slow component. In [6], a similar competition was observed in YAP:Ce SC between the processes of defects creation and the excitation of the $Ce^{3+}$ emission in the exciton absorption region. As the absorption/excitation bands of $Ce^{3+}$ are strongly overlapped with the intrinsic emission bands of YAP SC and SCF (see Fig. 1), one can assume that the appearance of the slow components in the $Ce^{3+}$ emission decay is caused by the energy transfer from the corresponding defects to $Ce^{3+}$ ions. However, the excitation bands of the defect-related emissions (Figs. 2b-5) do not coincide with the excitation spectrum of the slow component of the $Ce^{3+}$-related emission in the exciton absorption region (Fig. 9). Only in the 7.6-7.7 eV range, the slow component of the $Ce^{3+}$ emission in YAP SC can arise from the above-mentioned energy transfer processes. Therefore, one can assume that the slow component of the $Ce^{3+}$ emission appears around 8.0 eV mainly due to the tunneling recombinations between electron centers and hole $Ce^{4+}$ centers produced as a result of the photostimulated disintegration of the regular exciton. The tunnelling recombination processes in YAP were considered also in [22, 31]. In the energy range <7.5 eV, the slow component can appear in YAP SC due to the tunneling recombinations in the pairs consisting of an antisite-defect-related electron center and hole $Ce^{4+}$ center, both produced as a result of the photoionization of the higher-energy excited states of $Ce^{3+}$ or the photostimulated electron transfer processes (see also [6]). The presence of {$Y^{3+}_{Al}$-defect} centers in YAP SC was confirmed by the ESR studies [23].

The comparison of the lowest-energy exciton bands in the excitation spectra of the lead-related emission (Fig. 3) and the $Ce^{3+}$ emission (Fig. 9b) indicates that there is no energy transfer from the lead-induced centers to $Ce^{3+}$-related centers. However, the energy transfer from $Ce^{3+}$ to dimer lead centers can take place (see also [10]).

### 3.5. $^{89}$Y and $^{27}$Al NMR study

We have also performed studies of $^{89}$Y, and $^{27}$Al NMR in YAP:Ce grinded single crystals and powders including also nanopowders and nanocrystalline ceramics with the aim to prove existence of $Y_{Al}$ and $Al_Y$ antisite defects and to determine possible secondary phases in these materials. Such defects are believed to act as effective traps for charge carriers. NMR provides a unique approach to quantitatively measure the site occupancies and other imperfections in crystalline and even amorphous or liquid materials. NMR spectra of powders also give structural information. However, in contrast to single crystal measurements such spectra need extensive simulations. In particular, we found that $^{89}$Y NMR is very sensitive to the presence of antisite-related defects while $^{27}$Al NMR is more suitable for detection of the unwanted secondary phases. $^{89}$Y NMR spectra measured in grinded single crystal and commercial powder are shown in Fig. 10. NMR spectrum in grinded single crystal is well fitted by orthorhombic symmetry chemical shift tensor with $\delta$ = -121 ppm and asymmetry parameter $\eta$ = 0.12. In commercial powder NMR spectrum is complex. It contains two components. One component is completely similar to that measured in grinded single crystal and the second one is described by the following parameters: $\delta$ = -35 ppm and $\eta$ = 0.62. The second component is probably related to both $Y_{Al}$ antisite-related defects and resonances from near-surface regions of crystallites. Calculation shown that the total concentration of such defective Y sites in commercial YAP powder is up to 15-20%. On other hand, in commercial YAP crystals only $Y_{Al}$ antisite defects were detected with the concentration one order lower, i.e. about 2-3% [23].

$^{27}$Al NMR spectra were measured in grinded single crystals, microcrystalline powder and nanocrystalline ceramics (Fig. 11). One can see that the spectra in grinded single crystal and microcrystalline powder are practically similar while the spectrum in the nanocrystalline ceramics is much broader. It contains several components and only one component is related to the perovskite phase of YAP. Calculation shows that the nanocrystalline ceramics studied by us contained about 80% of the secondary non-perovskite phase. Therefore the synthesis method of YAP nanopowder has to be essentially improved.

## 4. Conclusions

The detailed study of luminescence characteristics of undoped and $Ce^{3+}$-doped SC and SCF of YAP and the analysis of published data has allowed us to identify the origin and structure of the defects responsible for various localized exciton emission bands. In YAP SCF, the relatively weak ~5.6 eV emission of the STE is the dominating one. In YAP SC, the STE emission is practically absent. The 5.63 eV emission arises from $ex^0Y^{3+}_{Al}$, while the 4.12 eV emission, from $ex^0\{Y^{3+}_{Al}\text{-}V_O\}$. The antisite defects of the type of $Y^{3+}_{Al}$ and $Y^{3+}_{Al}$-$V_O$ are absent in the YAP SCF, nanopowders and ceramics studied. In X-irradiated YAP:Ce SC, the thermally stimulated destruction of the corresponding electron centers around 200 K and 280 K, respectively, and recombination of the released electrons with $Ce^{4+}$ centers is accompanied with the $Ce^{3+}$ emission. The possibility of an effective formation of the {$Y^{3+}_{Al}$-$V_O$}-type defects in YAP single crystals is confirmed by the theoretical calculations.

In the excitation spectra of various intrinsic, defect-related and impurity-related emission bands, the regular and various perturbed exciton bands are identified. The comparison of their positions allowed to clarify the influence of various defects on the appearance of undesirable slow components in the luminescence decay kinetics of $Ce^{3+}$ centers which negatively influence the scintillation characteristics of YAP:Ce. The conclusion is made that the slow decay is mainly caused by the tunnelling recombination of electron and hole centers created at the disintegration of the regular exciton states and the photoionization of $Ce^{3+}$ ions. The contribution of the energy transfer processes between defect and $Ce^{3+}$ states is found to be smaller.

**Acknowledgements**

The work was supported by the projects of the Estonian (No. 7507), Czech (No. 202/08/0893 and GA AV IAA100100810), and Ukrainian (No. SF-28F) Science Foundations as well as by the II-20052049 EC and II-2009-0087 projects of DESY, Hamburg.

**Figure captions**

Fig. 1. (a) Emission spectra of YAP SCF measured under 8.4 eV (curve 1), 8.15 eV (curve 2), 7.95 eV (curve 3) and 7.6 eV (curve 4) excitations. (b) Time-resolved emission spectra of YAP SC measured under $E_{exc}$=7.95 eV. Excitation spectrum of the fast component of the $Ce^{3+}$ emission in YAP:Ce SC (dotted line). T=10 K.

Fig. 2. Time-resolved excitation spectra of YAP SCF (shown in the 6-9 eV range) measured at 10 K for (a) the STE emission and (b) the 4.16 eV emission of the localized exciton.

Fig. 3. Time-resolved excitation spectra of YAP SCF (shown in the 4-10 eV range) measured at 10 K for (a) the 3.63 eV emission of single $Pb^{2+}$-based centers and (b) the 3.15 eV emission of dimer $Pb^{2+}$ centers.

Fig. 4. Time-resolved excitation spectra of YAP SC measured at 10 K for (a) $E_{em}$=5.7 eV and (b) $E_{em}$=5.2 eV.

Fig. 5. Time-resolved excitation spectra of YAP SC (shown in the 6-9 eV energy range) measured for: (a) $E_{em}$=5.7 eV and (b) $E_{em}$=4.1 eV. T=10 K.

Fig. 6. (a) X-ray excited emission spectra of YAP single crystal, ceramics, and nanopowder measured at 80 K at the same conditions.

Fig. 7. TSL glow curves measured with the heating rate 5 K/min for different YAP:Ce samples after their X-ray irradiation at the same conditions (70 K, 30 min, 40 kV, 15 mA): (a) single crystal (7566/1) (solid line) and powder QM58/N-S1 (Phosphor Technology UK) (dashed line); (b) ceramics (2 wt% Ce) (solid line) and nanopowder (2 wt% Ce) (dashed line).

Fig. 8. The total energy dependence on the displacement of $Y^{3+}$ ion which replaces $Al^{3+}$ obtained for the case when the oxygen vacancy is close by.

Fig. 9. Time-resolved excitation spectra of the $Ce^{3+}$ emission measured at 10 K for (a) YAP:Ce SC and (b) YAP:Ce SCF (shown only in the 6-9 eV range).

Fig. 10. $^{89}Y$ NMR spectra measured at Larmor frequency 19.607 MHz in (a) grinded single crystal and (b) commercial powder of YAP. Points are measured spectra and solid lines are computer simulated spectra.

Fig. 11. $^{27}$Al NMR spectra measured in grinded single crystal, microcrystalline powder and nanocrystalline ceramics.

Table 1. Luminescence characteristics of perturbed and localized excitons and defects in YAP SC and YAP SCF at 10 K. The maxima positions obtained in the present paper at the same conditions are shown in bold.

| Sample | Exciton | Exciton emission, eV | Defect emission, eV | Lowest-energy exciton band, eV |
|---|---|---|---|---|
| YAP SC | $ex^0Y^{3+}_{Al}$ | 5.9 [12]; 5.69 [8]; **5.63** | - | 7.88 [8]; 7.95 [16]; **7.9** |
| | $ex^0\{Y^{3+}_{Al}-V_O\}$ | **4.12** | 2.45 [29] | **7.7; 7.13** |
| | $ex^0V_O$ | 4.2 [24]; | - | 7.67 [17]; 7.6 [13]; 7.52 [8] |
| | $ex^0Ce^{3+}$ | No | $Ce^{3+}$ | 7.80 [9]; **7.9** |
| | $ex^0\{Ce^{3+}-Y^{3+}_{Al}\}$ | No | ~$Ce^{3+}$ | 7.75 [9]; ~**7.7** |
| YAP SCF | STE | **~5.6** | **-** | 7.91 [17]; **7.83** |
| | $ex^0$ ($Pt^{4+}$) | **4.16** | **-** | **~7.6** |
| | $ex^0$ ($Pb^{2+}$) | 4.19 [8] | **-** | 7.59 [8] |
| | $ex^0$ ($Pb^{2+}$) | **No** | 3.63 [8] | **7.66** |
| | $ex^0$ ($Pb^{2+}$-$V_O$-$Pb^{2+}$} | No | 3.15 [10] | **7.74** |
| | $ex^0Ce^{3+}$ | No | $Ce^{3+}$ | 7.80 [9]; **7.9** |

Table 2. Structural parametres of $YAlO_3$ . Space group *Pmna* (53) and lattice parametres $a$ = 5.334 Å, $b$ = 7.375 Å, $c$ = 5,180 Å were taken from the x-ray diffraction data [27]. The fractional atomic coordinates, Y ($x$,1/4, $z$), Al (0,0,1/2) O1 ($x$,1/4,$z$), O2 ($x$, $y$, $z$) are obtained by total energy minimization and compared with x-ray refinement (experiment). There are four formula units per primitive cell.

| | Other code [28] | FPLO | Experiment [27] |
|---|---|---|---|
| Y x | 0.0562 | 0.0540 | 0.0526 |
| Y z | 0.9868 | 0.9881 | 0.9896 |
| O1 x | 0.4756 | 0.4756 | 0.475 |
| O1 z | 0.0884 | 0.0884 | 0.086 |
| O2 x | 0.2949 | 0.2949 | 0.293 |
| O2 y | 0.0465 | 0.0465 | 0.044 |
| O2 z | 0.7045 | 0.7045 | 0.703 |

Table 3. Calculated LDA total and defect energies of $YAlO_3$ based on 20-atom supercells.

| Type of defects | Energy of antisite pair defect (eV) | Total energy (Hartree) 1Hr=27.2 eV |
|---|---|---|
| $YAlO_3$ | | -15394.773028 |
| $Al_Y$-$Y_{Al}$ (antisite pair) | 10.8 | -15394.370739 |
| $Al_Y$-$Y_{Al}$ ($V_O$ is nearby) | 8.1 | -15319.478418 |
| $Al_Y$-$Y_{Al}$ ($V_O$ is far) | 10.6 | -15319.384259 |
| $Al_YY_{Al}$ – $Ce^{3+}{}_Y$ | 10.2 | -20871.244242 |
| $Y_{Al}$ ($V_O$ is far) | | -18459.58375 |
| $Y_{Al}$ ($V_O$ is nearby) | | -18459.66158 |
| $Ce^{3+}{}_Y$ ($V_O$ is nearby) | | -20796.646814 |
| $Ce^{3+}{}_Y$ ($V_O$ is far) | | -20796.637682 |

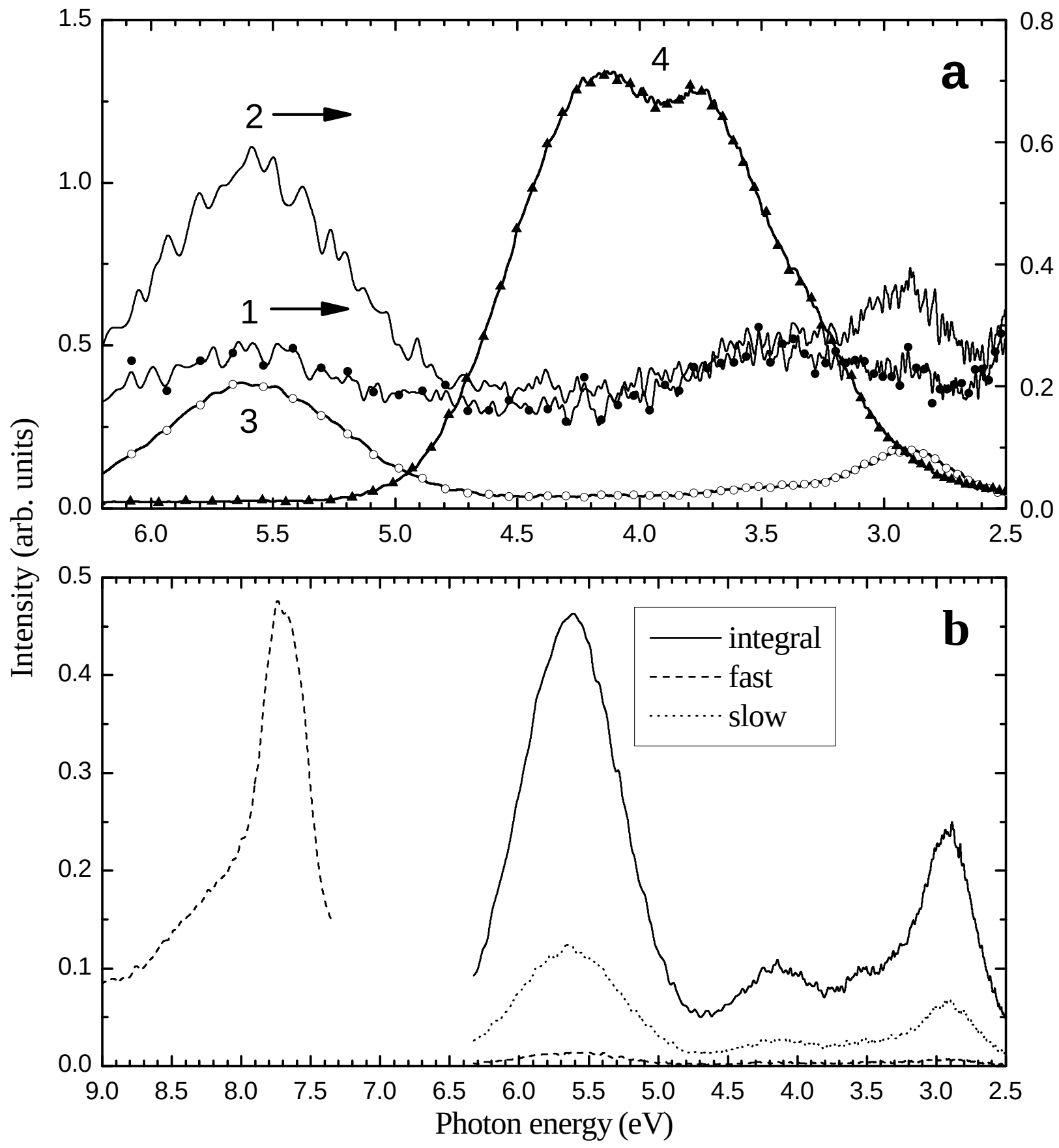

Fig. 1. (a) Emission spectra of YAP SCF measured under 8.4 eV (curve 1), 8.15 eV (curve 2), 7.95 eV (curve 3) and 7.6 eV (curve 4) excitations. (b) Time-resolved emission spectra of YAP SC measured under $E_{exc}$=7.95 eV. Excitation spectrum of the fast component of the $Ce^{3+}$ emission in YAP:Ce SC (dotted line). T=10 K.

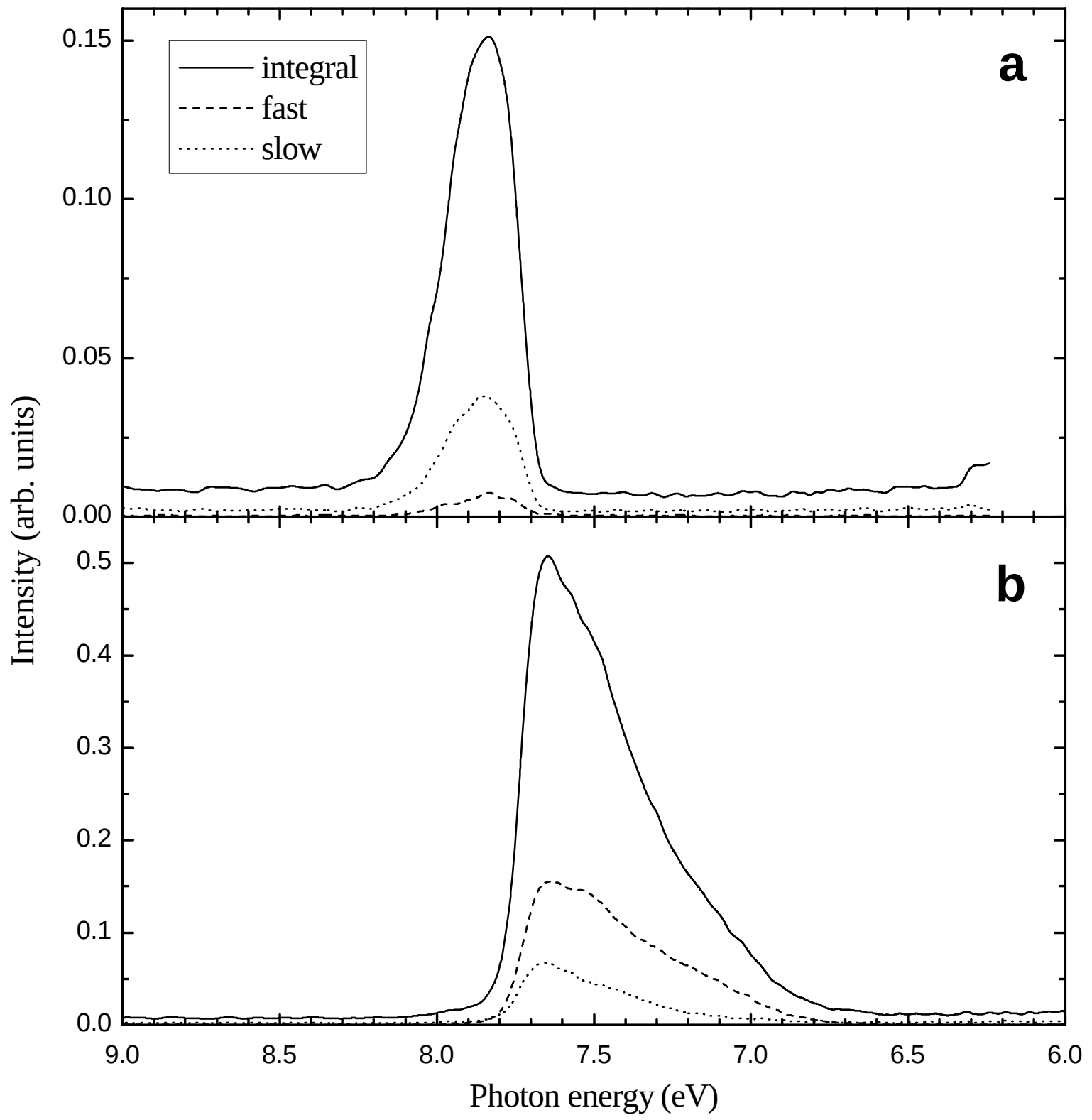

Fig. 2. Time-resolved excitation spectra of YAP SCF (shown in the 6-9 eV range) measured at 10 K for (a) the STE emission and (b) the 4.19 eV emission of the localized exciton.

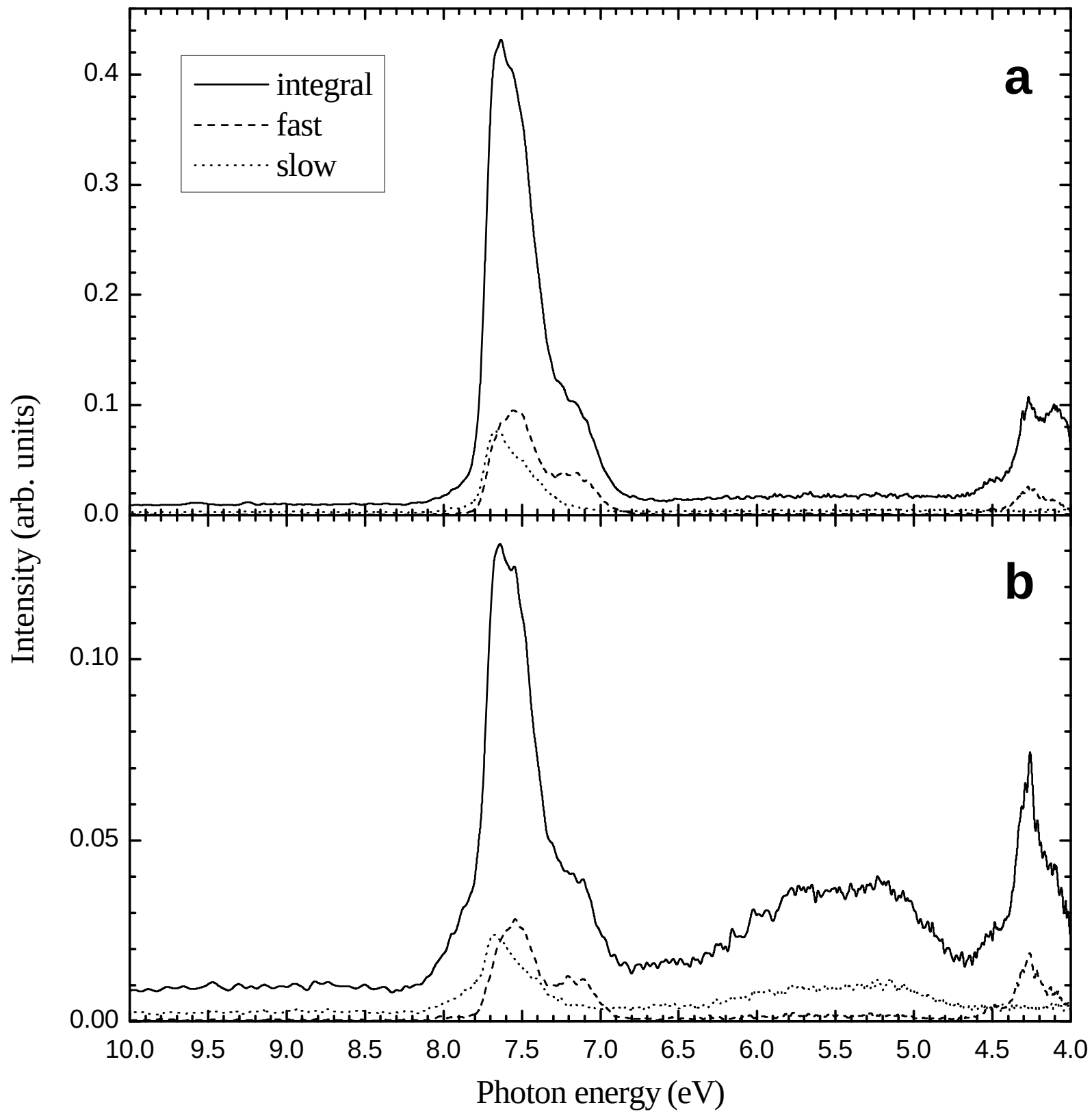

Fig. 3. Time-resolved excitation spectra of YAP SCF (shown in the 4-10 eV range) measured at 10 K for (a) the 3.63 eV emission of single $Pb^{2+}$-based centers and (b) the 3.15 eV emission of dimer $Pb^{2+}$ centers.

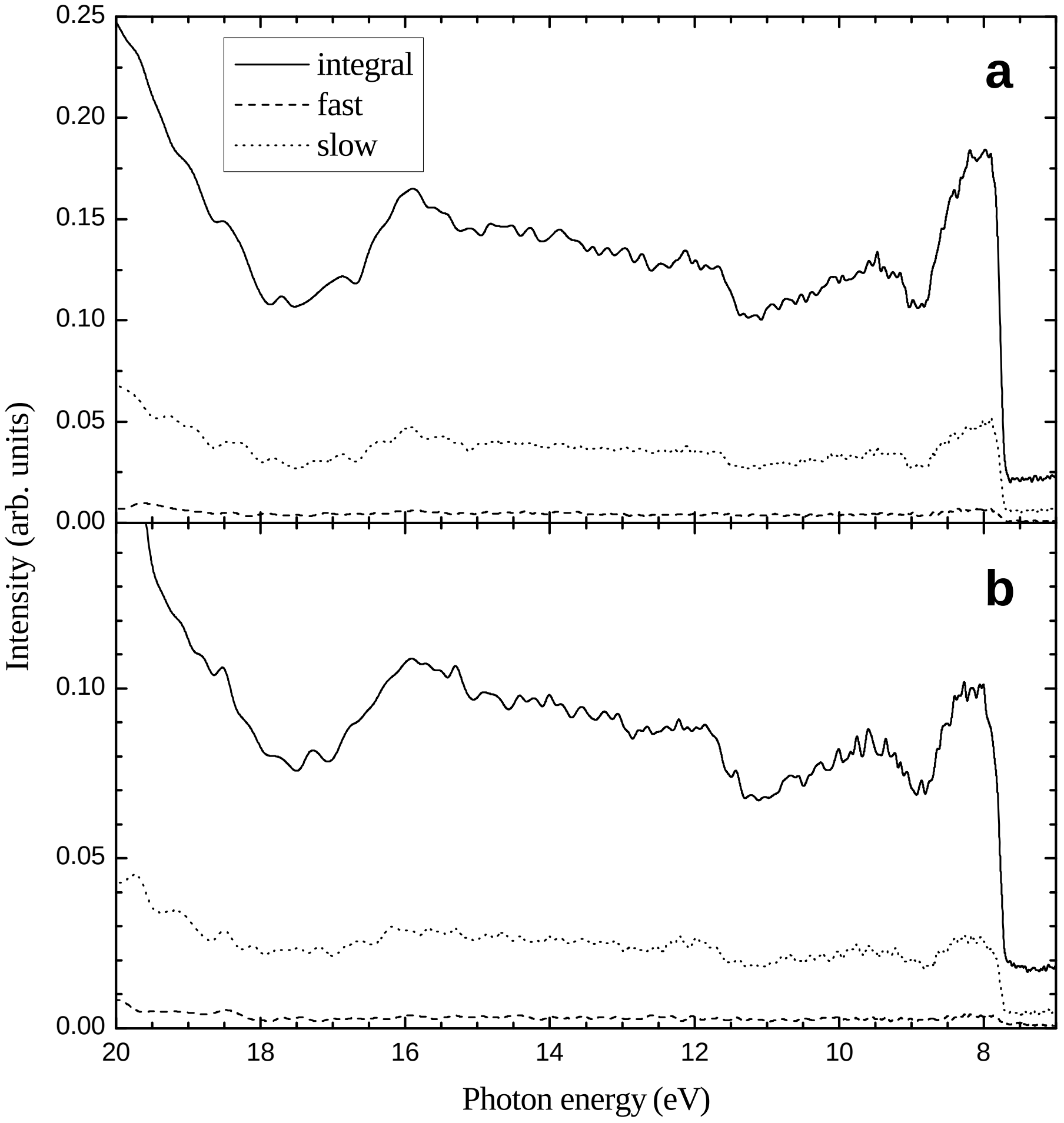

Fig. 4. Time-resolved excitation spectra of YAP SC measured at 10 K for (a) $E_{em}$=5.7 eV and (b) $E_{em}$=5.2 eV.

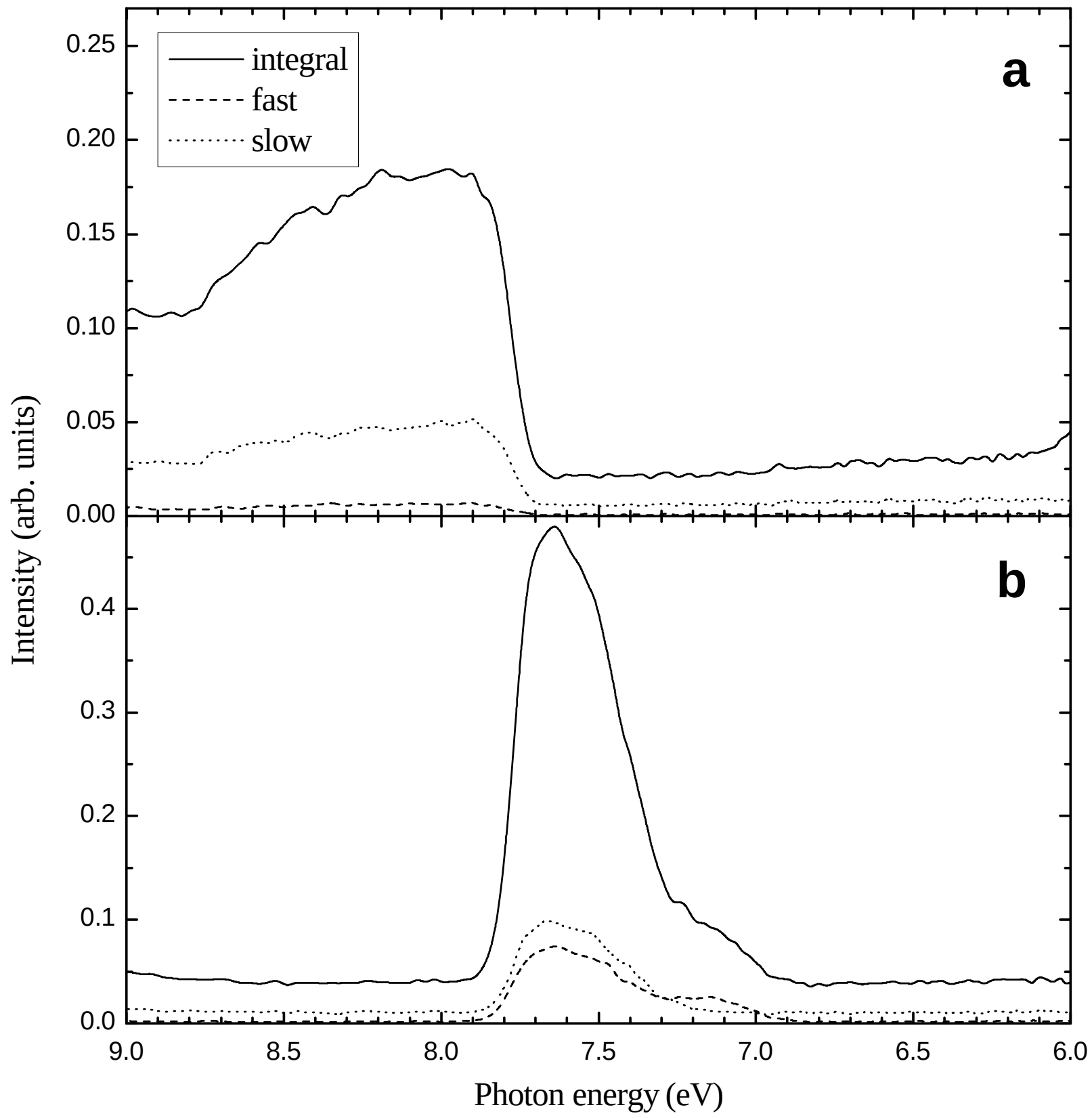

Fig. 5. Time-resolved excitation spectra of YAP SC (shown in the 6-9 eV energy range) measured for: (a) $E_{em}$=5.7 eV and (b) $E_{em}$=4.1 eV. T=10 K.

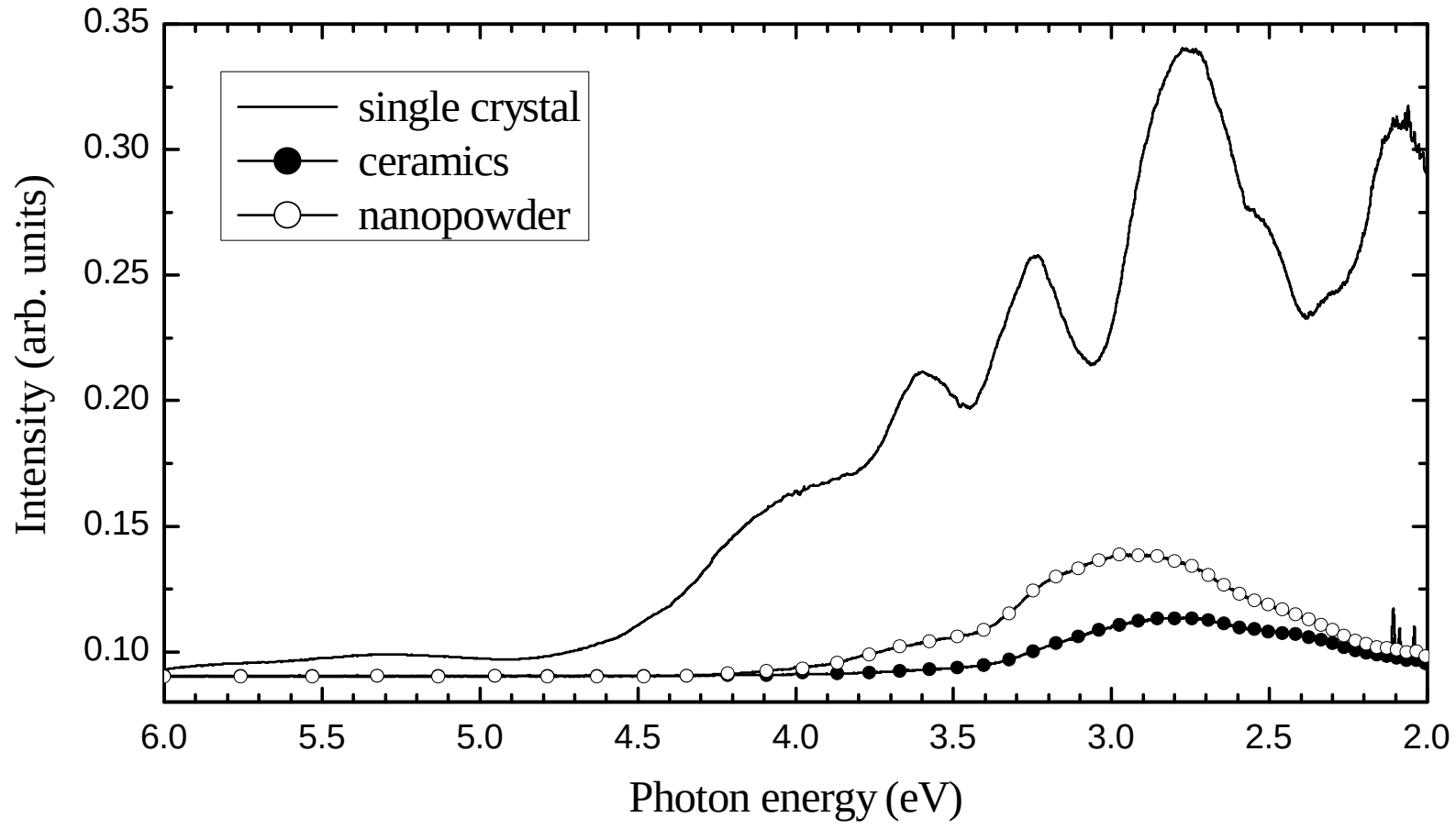

Fig. 6. (a) Emission spectra of YAP single crystal, ceramics and nanopowder measured at the same conditions at 80 K under the X-ray excitation.

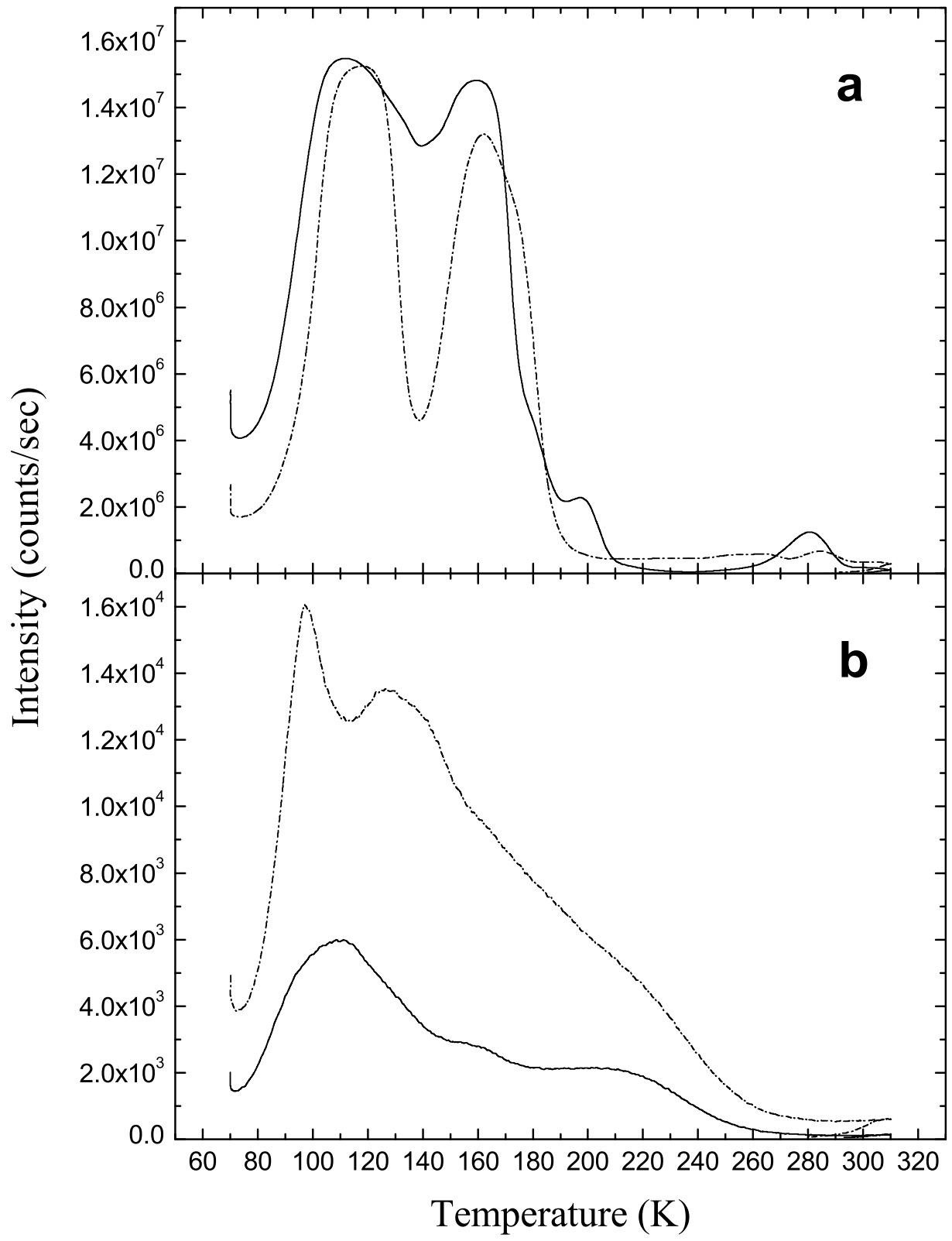

Fig. 7. TSL glow curves measured with the heating rate 5 K/min for different YAP:Ce samples after their X-ray irradiation at the same conditions (70 K, 30 min, 40 kV, 15 mA): (a) single crystal (7566/1) (solid line) and powder QM58/N-S1 (Phosphor Technology UK) (dashed line); (b) ceramics (2 wt% Ce) (solid line) and nanopowder (2 wt% Ce) (dashed line).

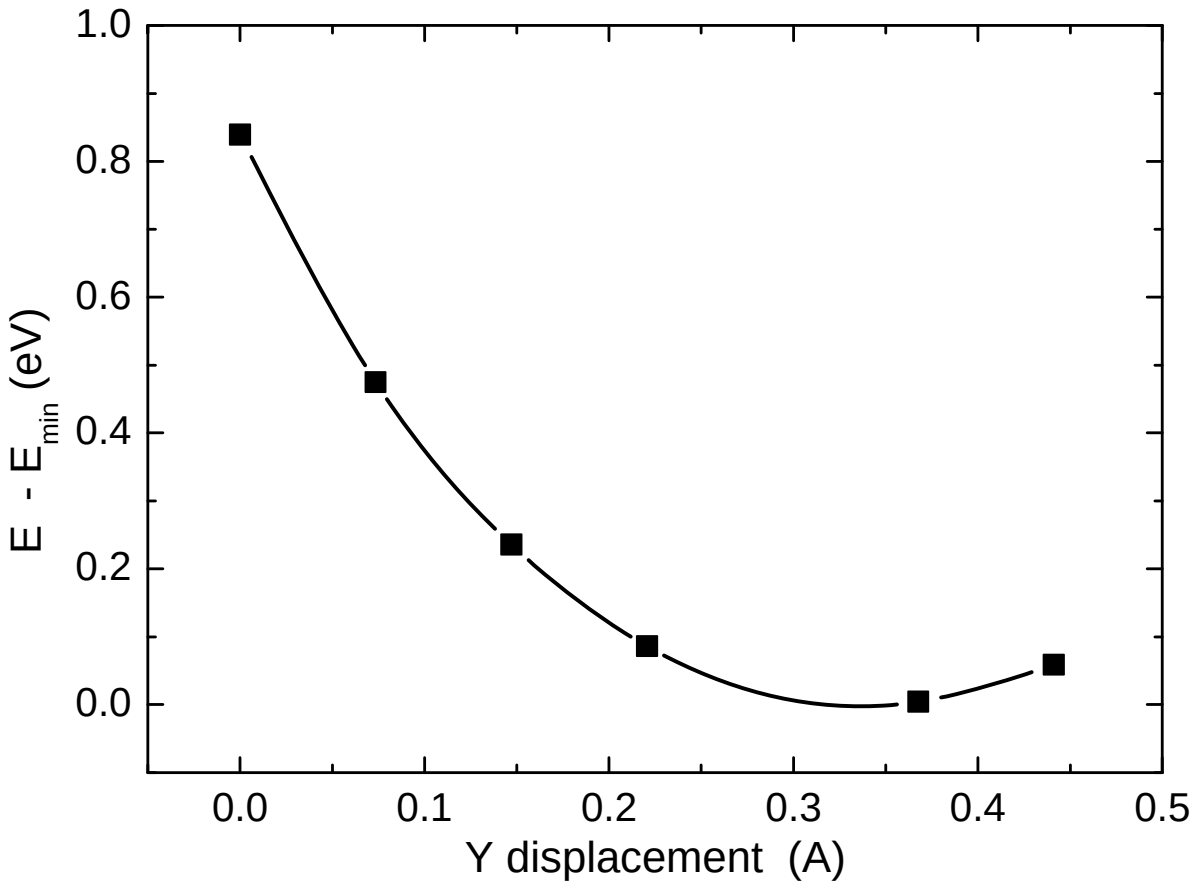

Fig. 8. The total energy dependence on the displacement of $Y^{3+}$ ion which replaces $Al^{3+}$ obtained for the case when the oxygen vacancy is close by.

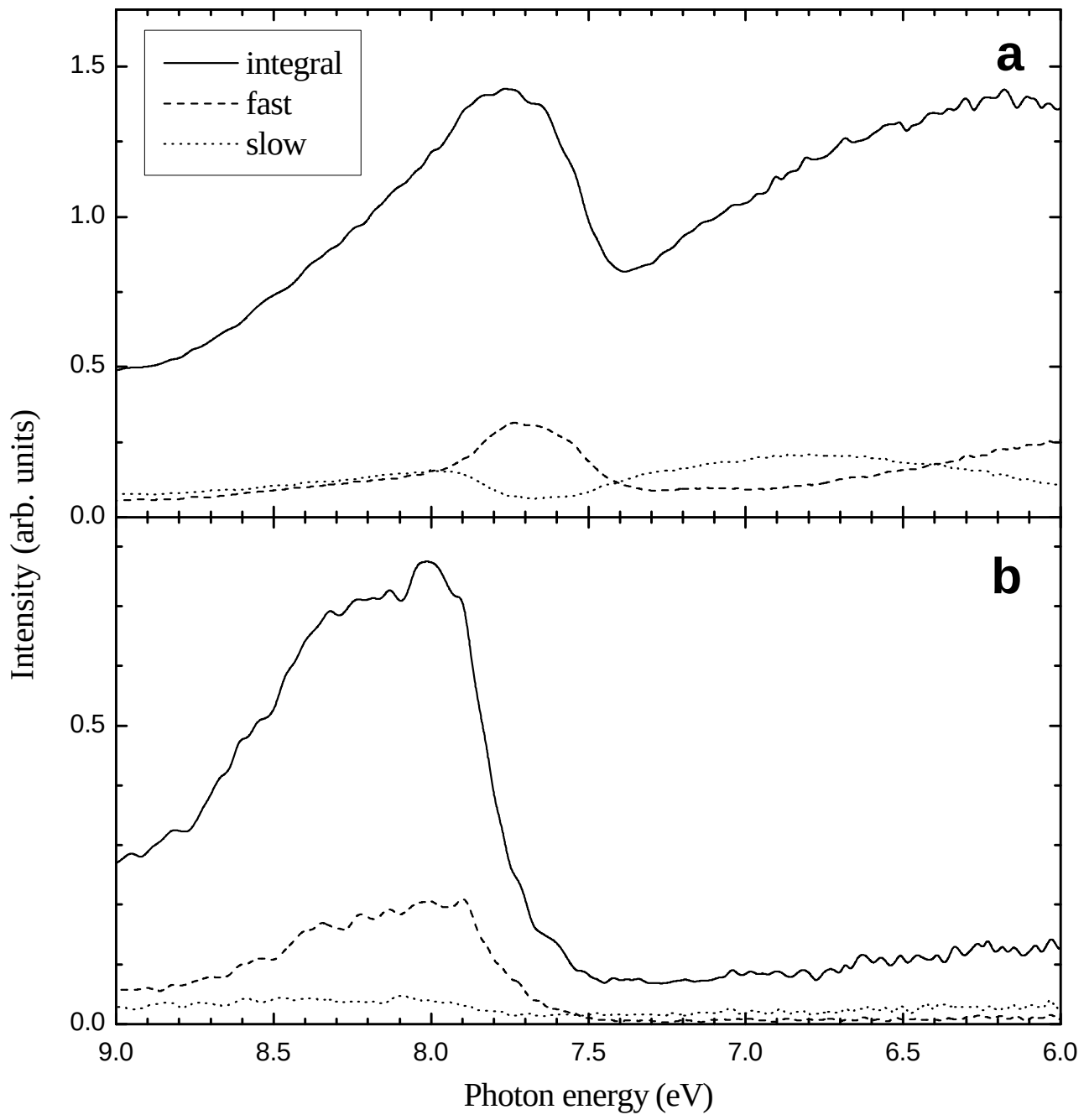

Fig. 9. Time-resolved excitation spectra of the $Ce^{3+}$ emission measured at 10 K for (a) YAP:Ce SC and (b) YAP:Ce SCF (shown only in the 6-9 eV range).

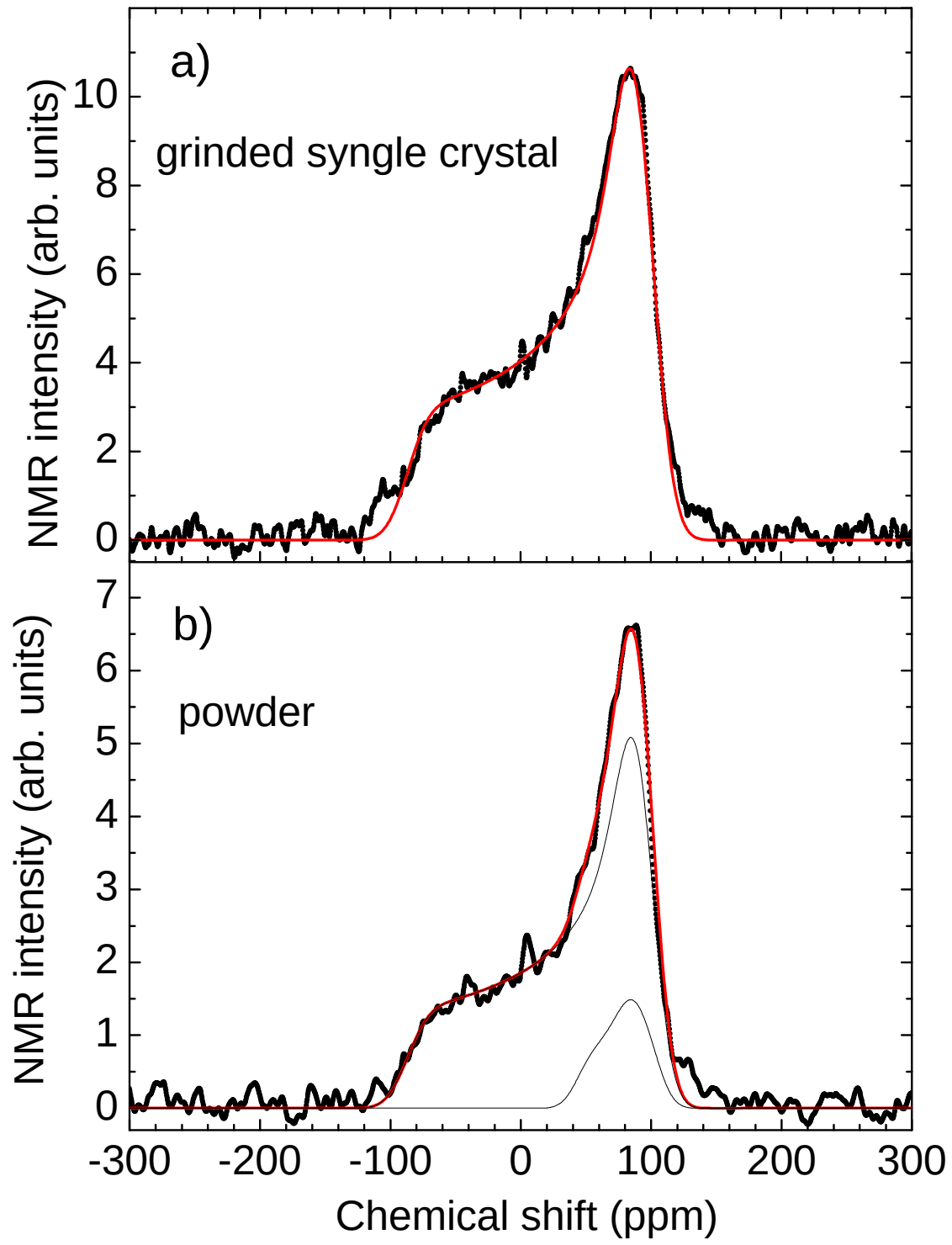

Fig. 10. $^{89}Y$ NMR spectra measured at Larmor frequency 19.607 MHz in (a) grinded single crystal and (b) commercial powder of YAP. Points are measured spectra and solid lines are computer simulated spectra.

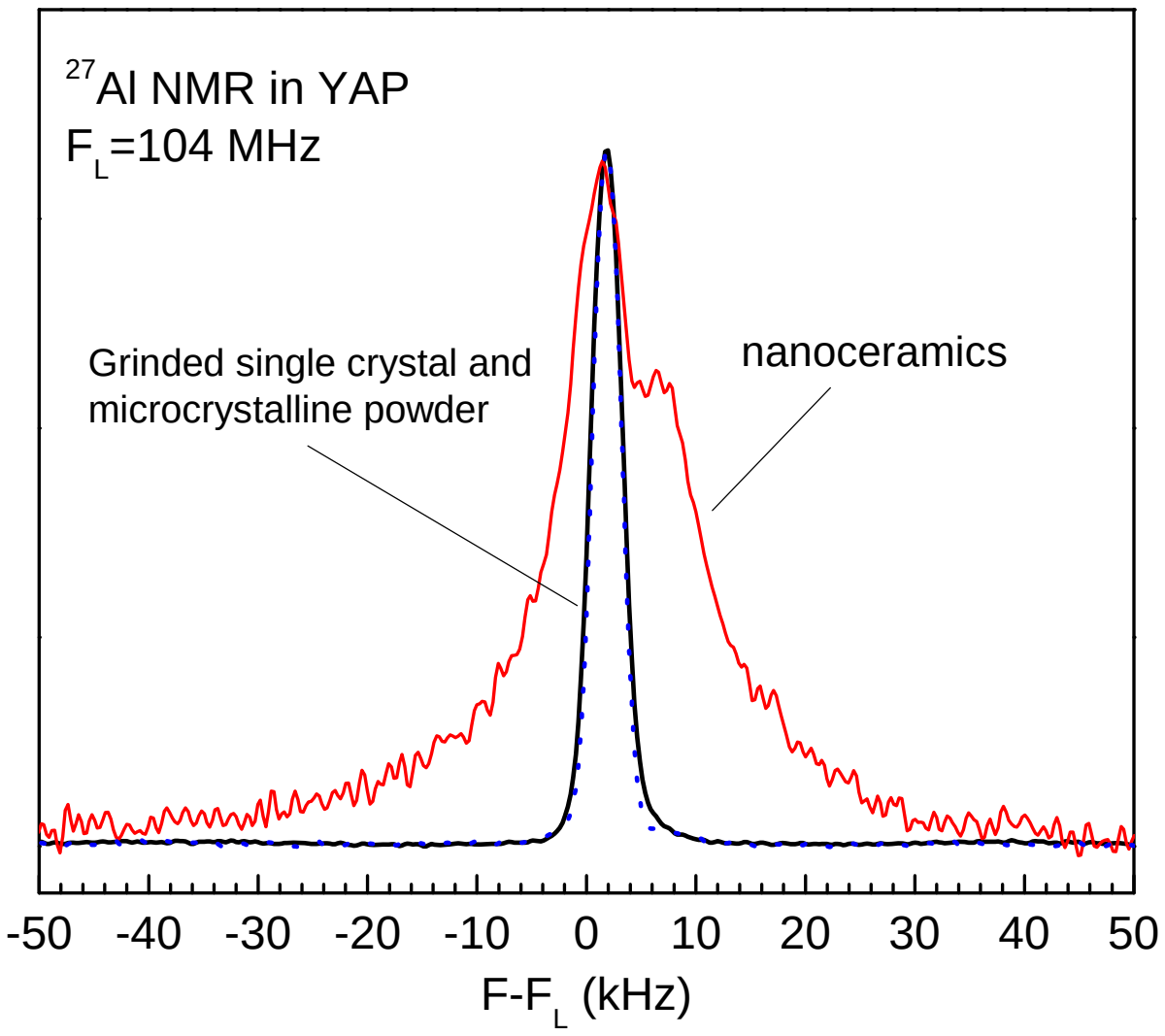

Fig. 11. $^{27}$Al NMR spectra measured in grinded single crystal, microcrystalline powder and nanocrystalline ceramics.